\documentclass[12pt]{iopart}

\usepackage{iopams}
\usepackage{graphicx}  % needed for figures
\usepackage{subfig} % to place pictures next to each other

\newcommand{\ket}[1]{\left|#1\right\rangle}

\begin{document}

\title[Robust states]{Robust entangled qutrit states in atmospheric turbulence}
\author{Tobias Br\"unner$^1$ and Filippus S Roux$^2$}
\address{$^1$ Institute of Physics, Albert-Ludwigs University of Freiburg, Hermann-Herder Str.\ 3, 79104 Freiburg, Germany}
\address{$^2$ CSIR National Laser Centre, P.O. Box 395, Pretoria 0001, South Africa}
\ead{fsroux@csir.co.za}

\begin{abstract}
The entangled quantum state of a photon pair propagating through atmospheric turbulence suffers decay of entanglement due to the scintillation it experiences. Here we investigate the robustness against this decay for different qutrit states. We use an infinitesimal-propagation equation to obtain the density matrix as a function of the propagation distance and we use the tangle to quantify the entanglement between a pair of qutrits. The evolution of various initial states as they propagate through turbulence is considered. Using optimization of the initial parameters, we obtain expressions for bipartite qutrit states that retain their initial entanglement longer than the initially maximally entangled states.
\end{abstract}


\submitto{\NJP}

%\maketitle

\section{Introduction}

Free-space quantum communication is one of the major components in the new quantum information technology revolution. A challenge that confronts free-space quantum communication is the decay of entanglement that entangled photons experience due to scintillation while propagating through a turbulent atmosphere. The approaches to overcome this challenge are either based on methods to correct the output optical field, such as using adaptive optics \cite{ao}, or on choosing an input optical field that is to some extent robust against entanglement decay in atmospheric turbulence. Here we consider the latter approach. In other words, we investigate to what extent a quantum state can be optimized so that it will retain as much of its initial entanglement as possible, while propagating through turbulence. We'll refer to such an optimized quantum state as a robust state.

In this paper we consider bipartite qutrit states in the Laguerre-Gaussian (LG) modal basis, which we restrict to the three elements that have a radial modal index of zero ($p=0$) and azimuthal modal indices of $\ell=1,0,-1$, which gives a three-dimensional Hilbert space per photon ${\cal H}_3$. The orbital angular momentum (OAM) associated with an LG mode is proportional to $\ell$. If, instead of choosing $|\ell|\leq 1$, we use larger values of $\ell$ (higher OAM), the entanglement is expected to last longer \cite{qturb1}.  This follows from the observation that the coupling from one OAM mode into another is stronger the closer the two OAM values are to each other \cite{paterson,tyler}.

Here the infinitesimal-propagation equation (IPE) \cite{ipe} is used to calculate the $z$-dependence of the corresponding density matrix. We assume that the two photons propagate through different uncorrelated  regions of atmospheric turbulence. The turbulence is modeled by the Kolmogorov power spectral density \cite{scintbook}.

We quantify the entanglement between the two photons by the tangle $\tau$ \cite{mintert2,mintert3} as a function of $z$. The tangle is equal to the square of the concurrence for pure states and gives a lower bound for the square of the concurrence of mixed quantum states. To find the most robust qutrit states, we optimize the tangle at a given propagation distance $z>0$. The result is expressed in terms of the parameters for the initial pure state that will give the maximum tangle at that propagation distance.

It has been shown that if only one partie of a bipartite (or multipartite) state passes through a dissipative channel, the final amount of entanglement is proportional to the initial entanglement of the state \cite{konrad1,tiersch1,tiersch2,gour}. For such cases the initially maximally entangled states are also the most robust states. This argument has been extended to the case where all parties of an entangled state pass through dissipative channels, provided that these channels are trace preserving \cite{gheorghiu}. However, if both parties of an entangled bipartite state pass through uncorrelated dissipative channels that are not trace preserving, only an upper bound exists for the evolution of the entanglement \cite{konrad1,tiersch1,tiersch2}. Here we'll show in particular that for two photons propagating through a turbulent atmosphere, corresponding to two qutrit states passing through uncorrelated dissipative channels that are not trace preserving, the initially maximally entangled state is not also the most robust state.

\section{Initial state} 

The most general pure bipartite qutrit state, defined in terms of our Hilbert space ${\cal H}_3 \otimes {\cal H}_3$ is
\begin{equation}
|\psi\rangle = \sum_{m,n} c_{m,n} |m\rangle |n\rangle .
\label{eq:init}
\end{equation}
where $c_{m,n}$ represents complex coefficients and $|m\rangle$ and $|n\rangle$ are the OAM eigenstates for the respective subsystems, with $m,n\in\{1,0,-1\}$. The normalization of the initial state implies that
\begin{equation}
\sum_{m,n} |c_{m,n} |^2 = 1 .
\label{innorm}
\end{equation} 
We use a parameterization that implicitly obeys this normalization condition. It also removes an overall phase factor and incorporates the symmetry associated with an interchange of the $\ell=1$ and $\ell=-1$ states. This parameterization is given by
\begin{equation}
\eqalign{c_{1,1} = \sin(k_{\rm a})\sin(k_{\rm b})\sin(k_{\rm d})\sin(k_{\rm h})\exp[\rmi(q_{\rm a} + q_{\rm b} + q_{\rm d} + q_{\rm d})] \cr
c_{1,-1} = \cos(k_{\rm a})\sin(k_{\rm b})\sin(k_{\rm d})\sin(k_{\rm h})\exp[\rmi(-q_{\rm a} + q_{\rm b} + q_{\rm d} + q_{\rm h})] \cr
c_{-1,1} = \sin(k_{\rm c})\cos(k_{\rm b})\sin(k_{\rm d})\sin(k_{\rm h})\exp[\rmi(q_{\rm c} - q_{\rm b} + q_{\rm d} + q_{\rm h})] \cr
c_{-1,-1} = \cos(k_{\rm c})\cos(k_{\rm b})\sin(k_{\rm d})\sin(k_{\rm h})\exp[\rmi(-q_{\rm c} - q_{\rm b} + q_{\rm d} + q_{\rm h})] \cr
c_{0,0} = \cos(k_{\rm h})\exp(-\rmi q_{\rm h}) \cr
c_{1,0} = \sin(k_{\rm e})\sin(k_{\rm f})\cos(k_{\rm d})\sin(k_{\rm h})\exp[\rmi(q_{\rm e} + q_{\rm f} - q_{\rm d} + q_{\rm h})] \cr
c_{-1,0} = \cos(k_{\rm e})\sin(k_{\rm f})\cos(k_{\rm d})\sin(k_{\rm h})\exp[\rmi(-q_{\rm e} + q_{\rm f} - q_{\rm d} + q_{\rm h})] \cr
c_{0,1} = \sin(k_{\rm g})\cos(k_{\rm f})\cos(k_{\rm d})\sin(k_{\rm h})\exp[\rmi(q_{\rm g} - q_{\rm f} - q_{\rm d} + q_{\rm h})] \cr
c_{0,-1} = \cos(k_{\rm g})\cos(k_{\rm f})\cos(k_{\rm d})\sin(k_{\rm h})\exp[\rmi(-q_{\rm g} - q_{\rm f} - q_{\rm d} + q_{\rm h})] .}
\label{eq:parm}
\end{equation}
Due to the normalization condition and the removal of the overall phase we have 16 real parameters: 8 angles ($k_{\rm a}, ..., k_{\rm h}$) and 8 phases ($q_{\rm a}, ..., q_{\rm h}$).

The density matrix $\rho(z)$ is obtained as a function of the propagation distance by solving the IPE \cite{ipe} for an initial density matrix defined as $\rho(0)=|\psi\rangle \langle\psi|$, where $|\psi\rangle$ is given in (\ref{eq:init}), using the coefficients defined in (\ref{eq:parm}). Although, we assume that the initial state (\ref{eq:init}) is a pure state, the scintillation process causes it to become mixed during propagation, which necessitates a density matrix approach.

\section{The IPE}

The IPE \cite{ipe} is a set of coupled first order differential equations that describes the evolution of the density matrix for a biphoton in the OAM basis as a function of the propagation distance $z$ through a turbulent medium. It represents a multiple phase screen approach, as opposed to the single phase screen approach \cite{paterson}. As a result it can simulate both phase and intensity fluctuations, whereas the single phase screen approach can only simulate phase fluctuations.

In the IPE the turbulence model is specified in terms of a power spectral density. Here we'll use the Kolmogorov power spectral density \cite{scintbook}, given by 
\begin{equation}
\Phi(k) = 0.033\ C_{\rm n}^2 k^{-11/3} ,
\label{kolmog}
\end{equation}
where $C_{\rm n}^2$ is the refractive index structure constant, which quantifies the strength of the turbulence. In the solution of the density matrix, obtained from the IPE the strength of the turbulence is contained in a dimensionless parameter defined by 
\begin{equation}
\sigma = {\pi^{3/2} C_{\rm n}^2 w_0^{11/3} \over 6 \Gamma(2/3) \lambda^3} ,
\label{eq:sigma}
\end{equation}
where $w_0$ is the radius of the Gaussian envelope at the waist of the beam, $\lambda$ is the wavelength of the photons and $z_{\rm R}$ is the Rayleigh range, defined by
\begin{equation}
z_{\rm R} = {\pi w_0^2\over\lambda} .
\label{eq:zR}
\end{equation}

For the three-dimensional bipartite case the density matrix is a $9\times 9$ matrix. Therefore, the IPE for this density matrix is given as 81 first order differential equations that are divided into several sets of coupled equations and a few uncoupled equations. To solve the sets of coupled differential equations we use a perturbative approach, exploiting the fact that the terms that couple the different equations are multiplied by constants that are invariably an order of magnitude smaller that the uncoupled terms. 

After solving the differential equations to obtain the density matrix as a function of the propagation distance, one finds that its dependence on the propagation distance is governed by two functions that are defined by the following integrals
\begin{equation}
Z(t) \equiv \int_0^t (1 + \tau^2)^{5/6}\ \mathrm{d} \tau
\label{zfunk}
\end{equation}
and
\begin{equation}
H(t) \equiv \int_0^t (1 + \tau^2)^{5/6} \left( \frac{1 + \rmi\tau}{1 - \rmi\tau} \right)\ \mathrm{d} \tau ,
\label{hfunk}
\end{equation}
where we define a normalized propagation distance $t=z/z_{\rm R}$. For convenience we separate $H(t)$ into its real and imaginary parts $H(t) \equiv H_{\rm r}(t) + \rmi H_{\rm i}(t)$. These integrals can be solved to give expressions in terms of hypergeometric functions
\begin{eqnarray}
Z(t) = t\ _2{\rm F}_1 \left( \left[-\frac{5}{6},\frac{1}{2}\right],\left[\frac{5}{2}\right],-t^2\right) & \approx t \label{eq:Z} \\
H_{\rm r}(t) = \frac{11}{8} t\ _2{\rm F}_1 \left( \left[\frac{1}{6},\frac{1}{2}\right],\left[\frac{1}{2}\right],-t^2\right)-\frac{2}{8} \left(1+t^2\right)^{5/6} & \approx t \label{eq:Hr} \\
H_{\rm i}(t) = \frac{6}{5} \left[ (1+t^2)^{5/6}-1 \right] & \approx t^2 , \label{eq:Hi}
\end{eqnarray}
where $_2{\rm F}_1([\cdot],[\cdot],\cdot)$ represents Barnes's extended hypergeometric functions \cite{barnes}.

\begin{figure}[!ht]
\begin{center}
\subfloat[]{\scalebox{0.32}{\includegraphics{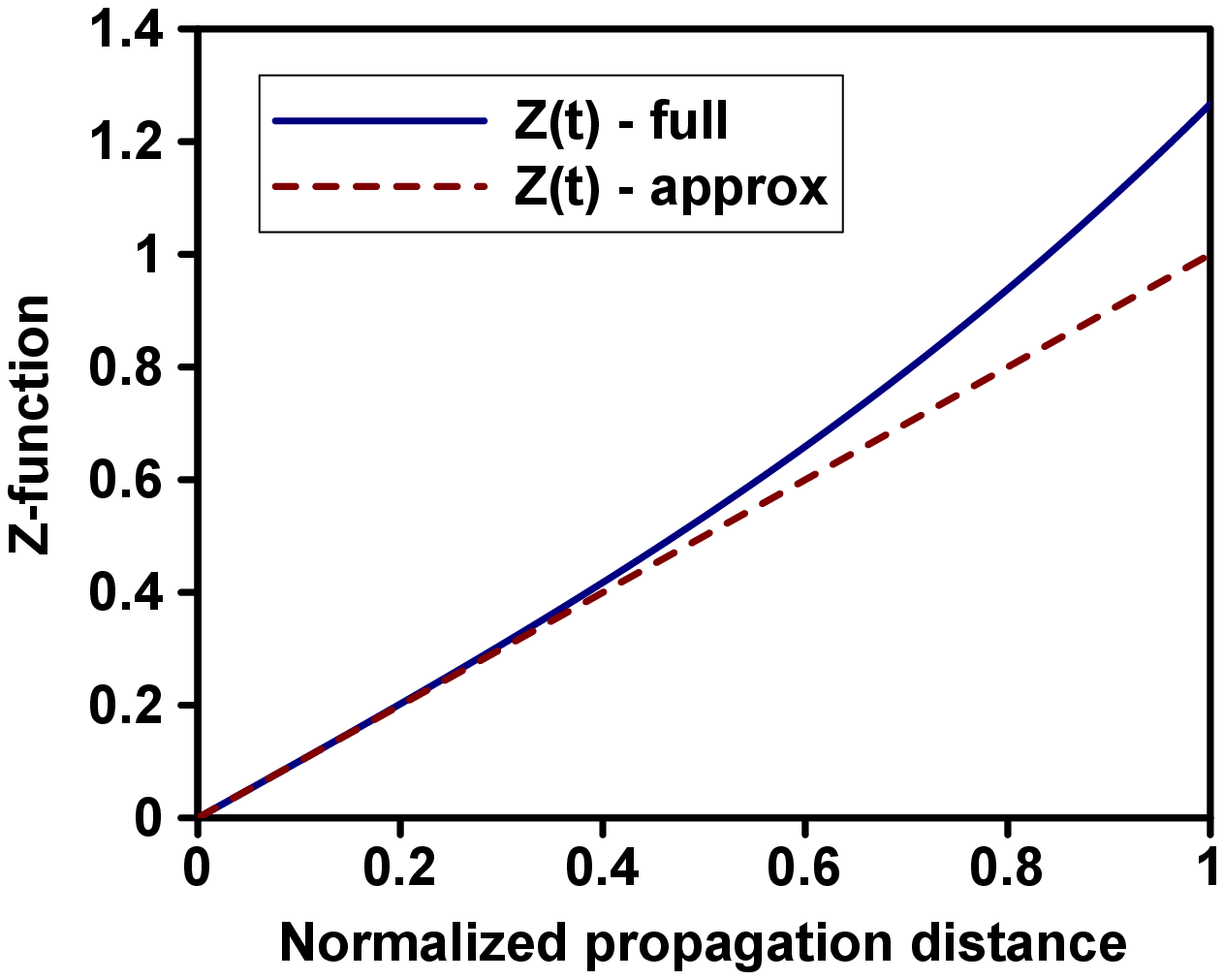}}\label{fig:z}}
\subfloat[]{\scalebox{0.32}{\includegraphics{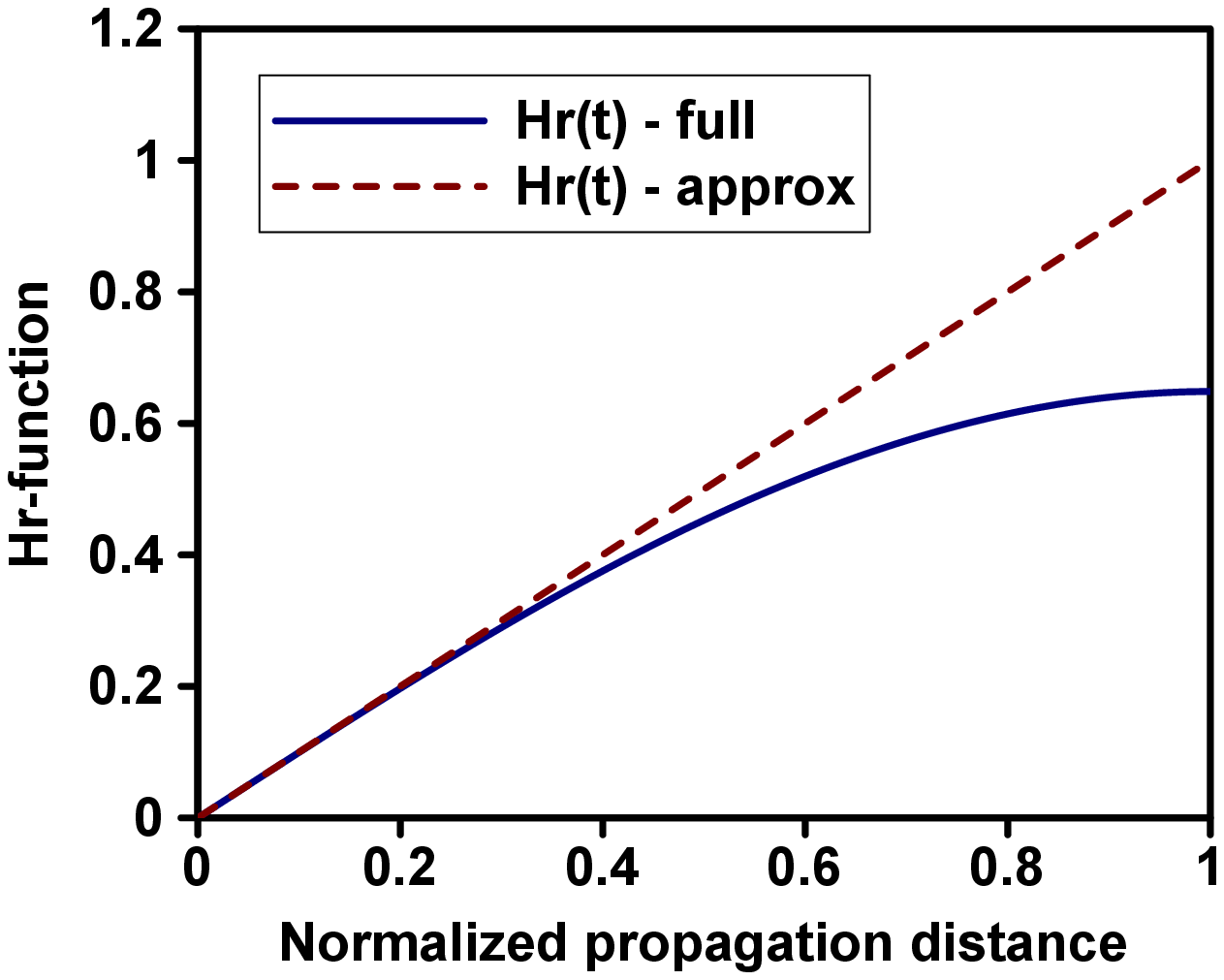}}\label{fig:hr}}
\subfloat[]{\scalebox{0.32}{\includegraphics{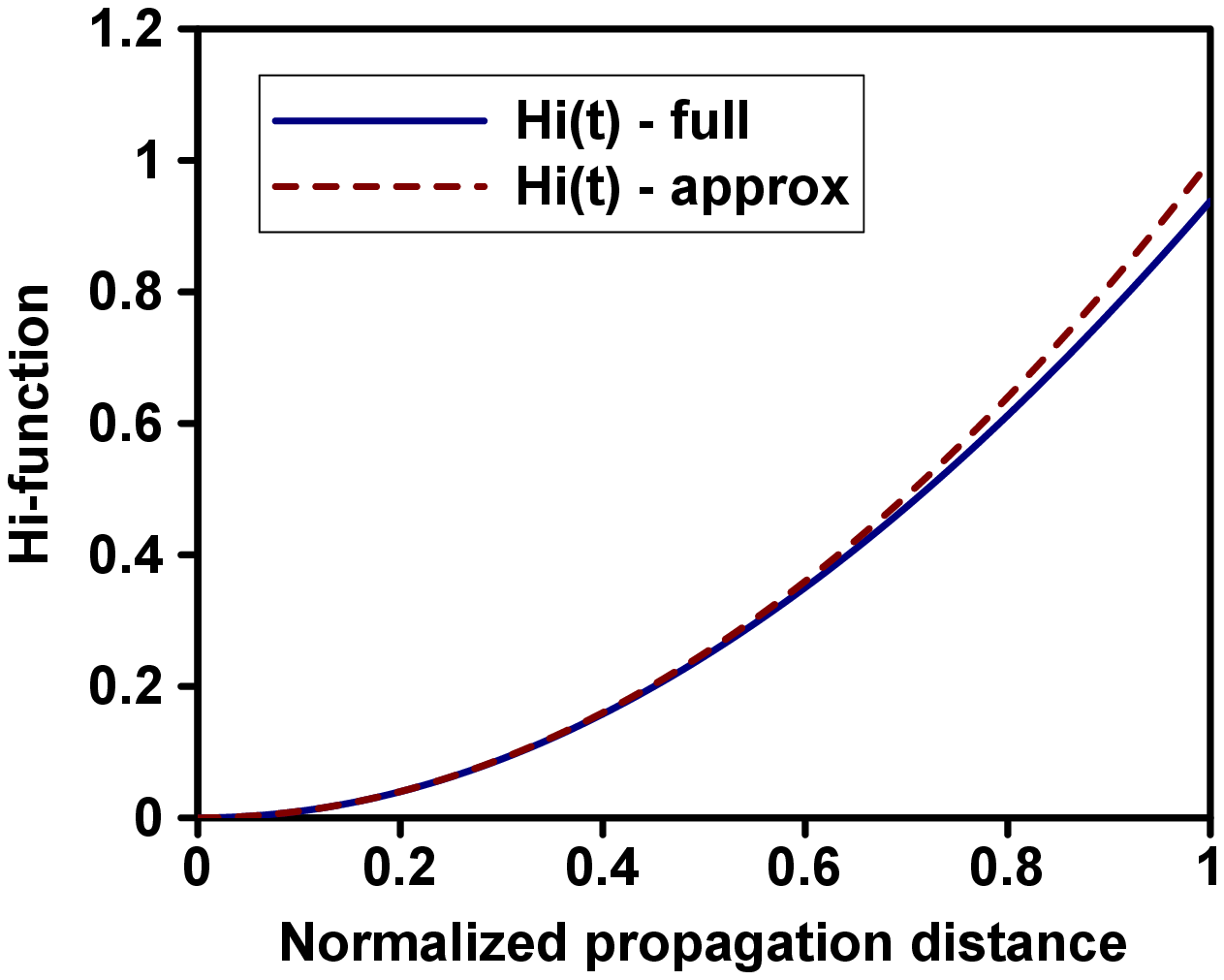}}\label{fig:hi}}
\end{center}
\caption{Full and approximate solutions of the integrals for $Z(t)$, $H_{\rm r}(t)$ and $H_{\rm i}(t)$ as a function of the normalized propagation distance $t$.} 
\label{fig:integrals}
\end{figure}

In figure~\ref{fig:integrals} we show the exact solutions for $Z(t)$, $H_{\rm r}(t)$ and $H_{\rm i}(t)$ in comparison with their approximations, indicating that if propagation is limited to $t \lesssim 1/3$, one can use the approximate solutions. The requirement that $t<1/3$ implies that the entanglement decays quickly, which calls for strong turbulence conditions. In what follows we'll consider both the cases where the entanglement decays at $t<1/3$ (strong turbulence), using the approximate expressions in (\ref{eq:Z}), (\ref{eq:Hr}) and (\ref{eq:Hi}), and the cases where the entanglement only decays at $t>1/3$ (weak turbulence), using the full expressions in (\ref{eq:Z}), (\ref{eq:Hr}) and (\ref{eq:Hi}).

\section{Trace}

The solution that is obtained from the IPE represents a truncated density matrix, because only three of the infinite number of basis elements are retained. This is experimentally analogous to post selection \cite{postselection} of measurement results in the three dimensional subspace. During propagation through turbulence the energy in the initial modes is scattered into higher order modes that are not contained in the truncated density matrix. As a result the density matrix is subnormalized --- its trace is smaller that 1. This, in turn, causes the trace of the truncated density matrix to decrease as a function of the propagation distance. Moreover, the function of the trace depends on the initial state. However, due to the symmetries among the chosen OAM basis elements, the trace function only contains the parameters $k_{\rm h}$ and $k_{\rm d}$.

\begin{figure}[!ht]
\begin{center}
\subfloat[]{\scalebox{0.48}{\includegraphics{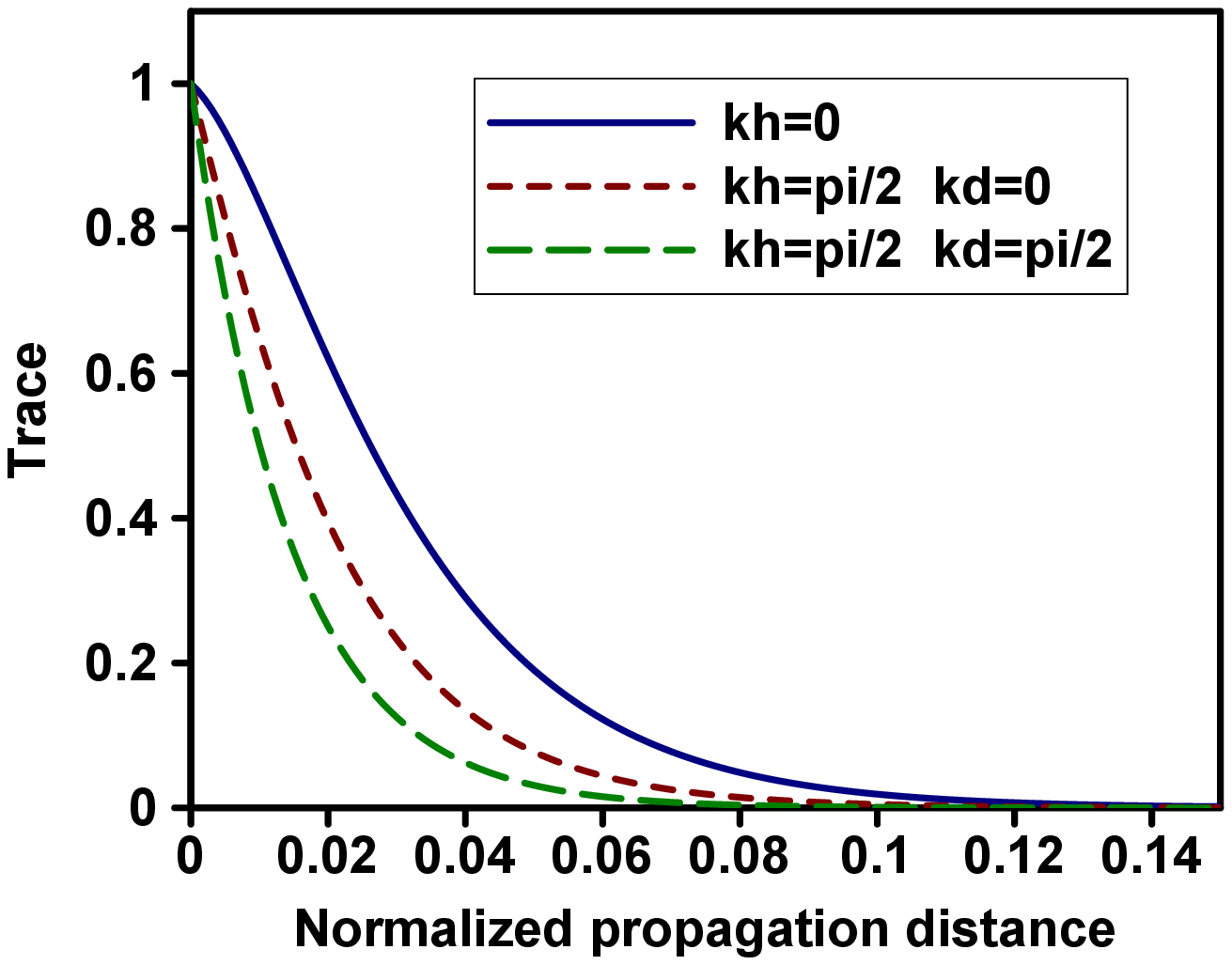}}\label{fig:trace1}}
\subfloat[]{\scalebox{0.48}{\includegraphics{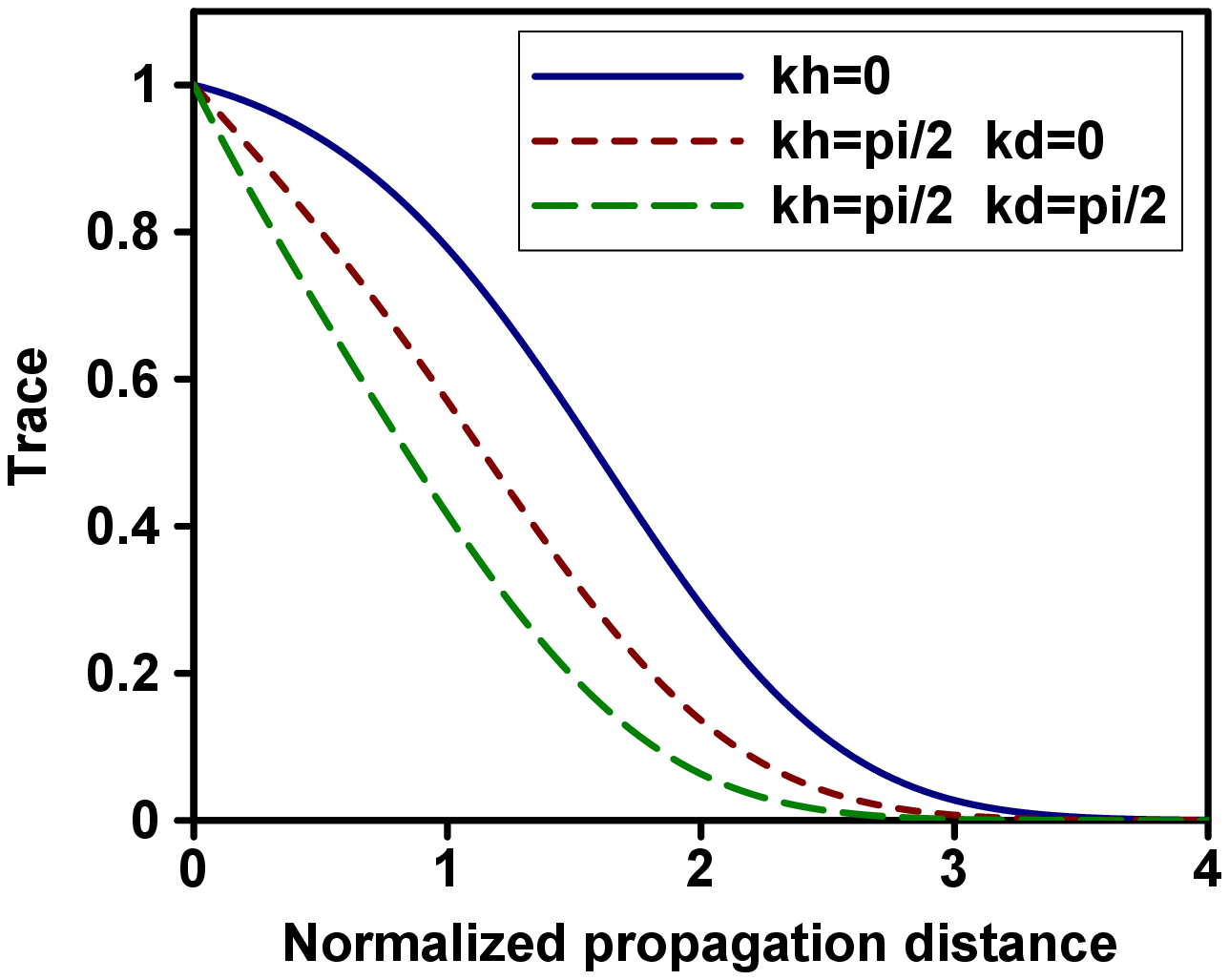}}\label{fig:trace2}}
\end{center}
\caption{Trace evolution of the truncated density matrix for strong turbulence (a) and for weak turbulence (b). Each graph contains three curves: solid (blue) line with $k_{\rm h} = 0$, short dashed (red) line with $k_{\rm h} = \pi/2,\ k_{\rm d} = 0$ and long dashed (green) line with $k_{\rm h} = k_{\rm d} = \pi/2$.} 
\label{fig:trace}
\end{figure}

The evolution of the trace as a function of $t$ is plotted in figure \ref{fig:trace} for three extreme cases:\ $k_{\rm h} = 0$, which is independent of $k_{\rm d}$; $k_{\rm h} = \pi/2,\ k_{\rm d} = 0$; and $k_{\rm h} = k_{\rm d} = \pi/2$. These cases are considered for strong turbulence ($\sigma = 25$) in figure \ref{fig:trace1} and for weak turbulence ($\sigma = 0.25$) in figure \ref{fig:trace2}. The plots show that states with $\ell=0$ scatter slower than those with $\ell=\pm 1$. In fact, the slowest decay of the trace is obtained for $k_{\rm h} = 0$, which represents a state consisting of only $|0,0\rangle$ and, therefore, is completely separable.

\section{Tangle}
\label{tangle}

For pure two-dimensional bipartite systems the concurrence \cite{wootters2} is a suitable measure of entanglement, but for mixed states in higher dimensions it is computationally demanding to calculate the concurrence directly through the construction of a convex roof \cite{mintert1}. Instead, we use the tangle $\tau\{\rho\}$, which is equal to the square of the concurrence for pure states and gives a lower bound for that of mixed quantum states \cite{mintert2}. The tangle is calculated via the purities of the (reduced) density matrices
\begin{equation}
\tau\{\rho\} = 2\ \tr\{\rho^2\} - \tr\{\rho_{\rm A}^2\} - \tr\{\rho_{\rm B}^2\} ,
\end{equation}
where $\rho_{\rm A}$ and $\rho_{\rm B}$ are the respective reduced density matrices of the two subsystems. For a maximally entangled state $\tau = \tau_{\rm max} = 2(d-1)/d$, where $d$ is the dimension of the subsystems' Hilbert spaces and for a separable state $\tau = \tau_{\rm min} = 0$.

\subsection{Bell states}

\begin{figure}[!ht]
\begin{center}
\subfloat[]{\scalebox{0.48}{\includegraphics{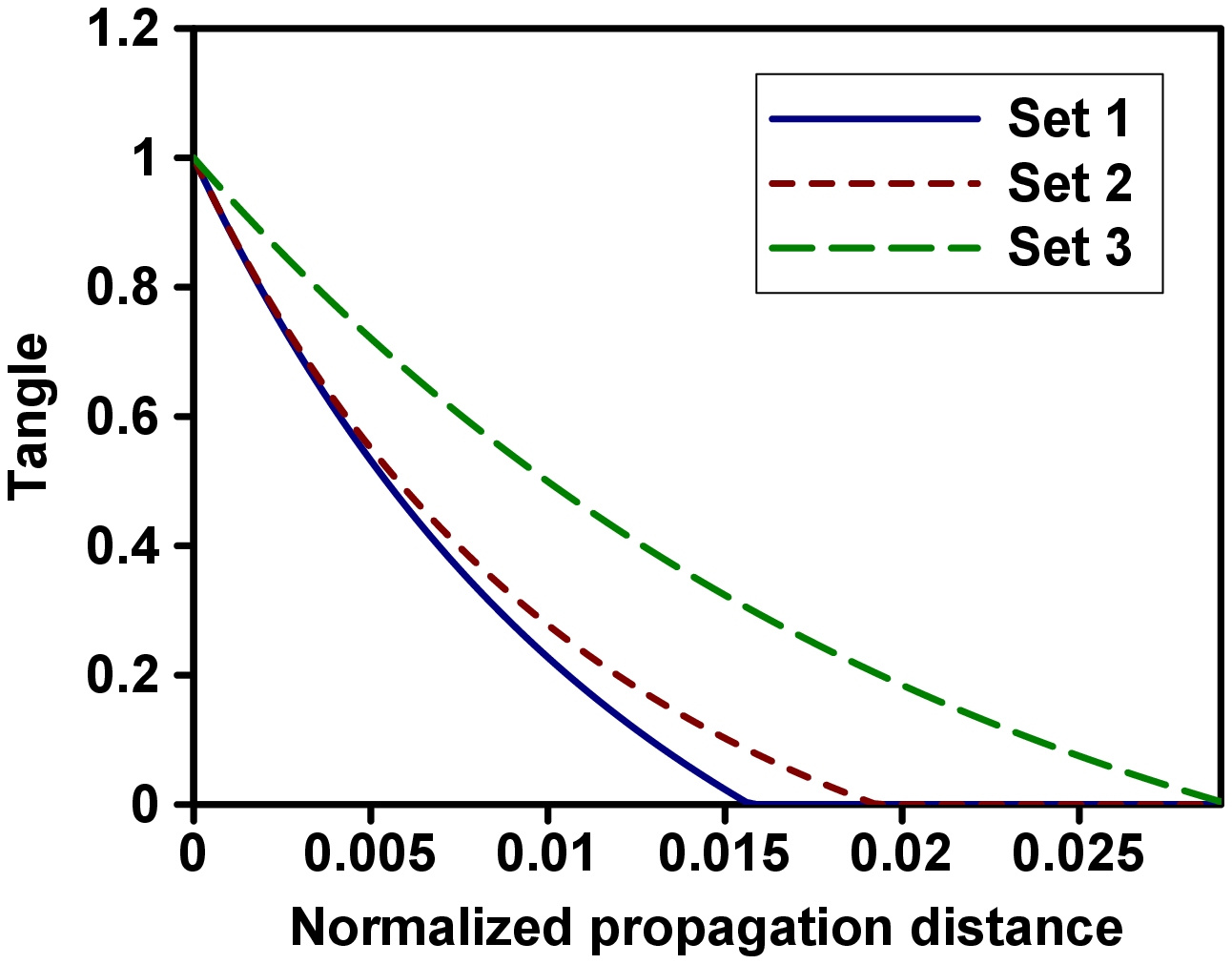}}\label{fig:Bell1}}
\subfloat[]{\scalebox{0.48}{\includegraphics{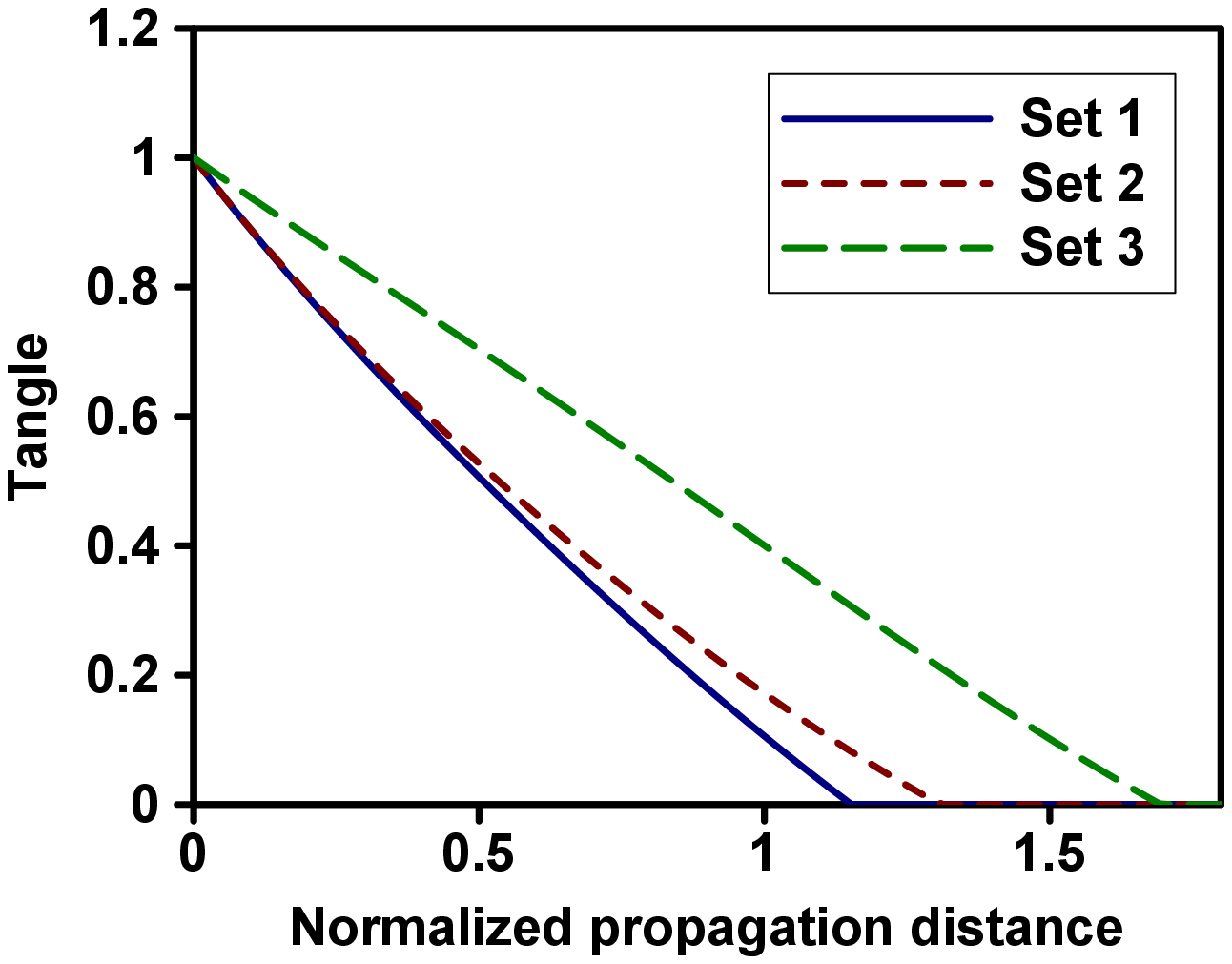}}\label{fig:Bell2}}
\end{center}
\caption{Comparison of the tangle evolution of the Bell states in Set 1, Set 2 and Set 3 for strong turbulence (a) and for weak turbulence (b).} 
\label{fig:Bell12}
\end{figure}

In the two-dimensional case, where diametric azimuthal indices $\ell=\pm 1$ are used, it can be shown that the entanglement of the four maximally entangled Bell states, quantified by the concurrence, all decay equally \cite{ipe}. This can be understood as a result of the fact that the scattering is symmetric with respect to $\ell = 1$ and $\ell = -1$. In the three-dimensional case there is a third possible azimuthal index, which does not share this symmetry. In our case we chose the third index to be $\ell = 0$, and we quantify the entanglement by the tangle. As a result, we find that the 12 Bell states in the two-dimensional subspaces of our three-dimensional Hilbert space form three sets based on their decay curves
\begin{eqnarray}
{\rm Set}~1 = \left \{ \ket{\Phi^+_{0,1}}, \ket{\Phi^+_{0,-1}}, \ket{\Phi^-_{0,1}}, \ket{\Phi^-_{0,-1}} \right \} \label{eq:set1} \\
{\rm Set}~2 = \left \{ \ket{\Psi^+_{0,1}}, \ket{\Psi^+_{0,-1}}, \ket{\Psi^-_{0,1}}, \ket{\Psi^-_{0,-1}} \right \} \label{eq:set2} \\
{\rm Set}~3 = \left \{ \ket{\Phi^+_{1,-1}},\ket{\Phi^-_{1,-1}},\ket{\Psi^+_{1,-1}},\ket{\Psi^-_{1,-1}} \right \} \label{eq:set3}
\end{eqnarray}
where 
\begin{eqnarray}
\ket{\Phi^{\pm}_{r,s}} = \frac{1}{\sqrt{2}}\left(\ket{r,r} \pm \ket{s,s} \right) \\
\ket{\Psi^{\pm}_{r,s}} = \frac{1}{\sqrt{2}}\left(\ket{r,s} \pm \ket{s,r} \right)
\end{eqnarray}
are the Bell states of the two-dimensional subspaces spanned by $r$ and $s$. The three sets of subspace Bell states give three different decay curves, as shown in figure \ref{fig:Bell12}.

Figure \ref{fig:Bell12} shows that the initial amount of entanglement does not uniquely determine the remaining entanglement after a propagation distance $t$, as was found for the two-dimensional situation \cite{ipe}. The Bell states of Set 3 maintain a non-zero entanglement for longer than the states of Sets 1 and 2. This can be understood as a consequence of the fact that there is a stronger coupling in this case between modes having azimuthal indices with different magnitudes than between those with identical magnitudes, which in turn follows from how the coupling strength between two OAM modes depends on the difference between their OAM-values.

\subsection{Initially maximally entangled states}

Next we investigate the decay of entanglement of initially maximally entangled qutrit states. Depending on their decay curves, one can identify two sets of initial qutrit states, given by
\begin{eqnarray}
{\rm State}~1 = \frac{1}{\sqrt{3}} \left[ \exp \left( \rmi \phi_1\right) \ket{j, k} + \exp \left( \rmi \phi_2 \right) \ket{-j, -k} + \exp \left( \rmi \phi_3\right) \ket{0, 0} \right] \label{eq:state1} \\
{\rm State}~2 = \frac{1}{\sqrt{3}} \left[ \exp \left( \rmi \psi_1\right) \ket{0, -j} + \exp \left( \rmi \psi_2 \right) \ket{-j, 0} + \exp \left( \rmi \psi_3\right) \ket{j, j} \right] , \label{eq:state2}
\end{eqnarray}
where $j, k \in \{-1, 1\}$, and $\phi_1$, $\phi_2$, $\phi_3$, $\psi_1$, $\psi_2$ and $\psi_3$ are arbitrary phases. Calculating the tangle for states with the initial conditions given as State 1 in (\ref{eq:state1}), we find that the phases always appear in the combination $\phi = \phi_1 + \phi_2 - 2 \phi_3$ in the expression of the tangle. The expression for the combined phase that maximizes the tangle at $t=t_0$ is given by
\begin{equation}
\phi = \phi_{\rm opt} \left(t_0\right) = \arctan \left[ \frac{2\ H_{\rm r}(t_0)\ H_{\rm i}(t_0)}{H_{\rm i}(t_0)^2 - H_{\rm r}(t_0)^2}\right] ,
\label{eq:phi}
\end{equation}
while $\phi = \phi_{\rm opt} \left(t_0\right) - \pi$ minimizes the tangle. Note that, although the combined phase that minimizes or maximizes the tangle depends on the propagation distance, this dependence is very small in the case of strong turbulence, because the entanglement decays quickly, for $t<1/3$. As a result we can use the small $t$ approximation, which implies that $\phi\approx 0$ and $\phi\approx\pi$, respectively, maximizes and minimizes the tangle. We'll denote the cases with minimum and maximum tangle as State 1a and State 1b, respectively. In the case where the initial state is given by State 2 the tangle is totally independent of the phases $\psi_1$, $\psi_2$ and $\psi_3$.

\begin{figure}[!ht]
\begin{center}
\subfloat[]{\scalebox{0.48}{\includegraphics{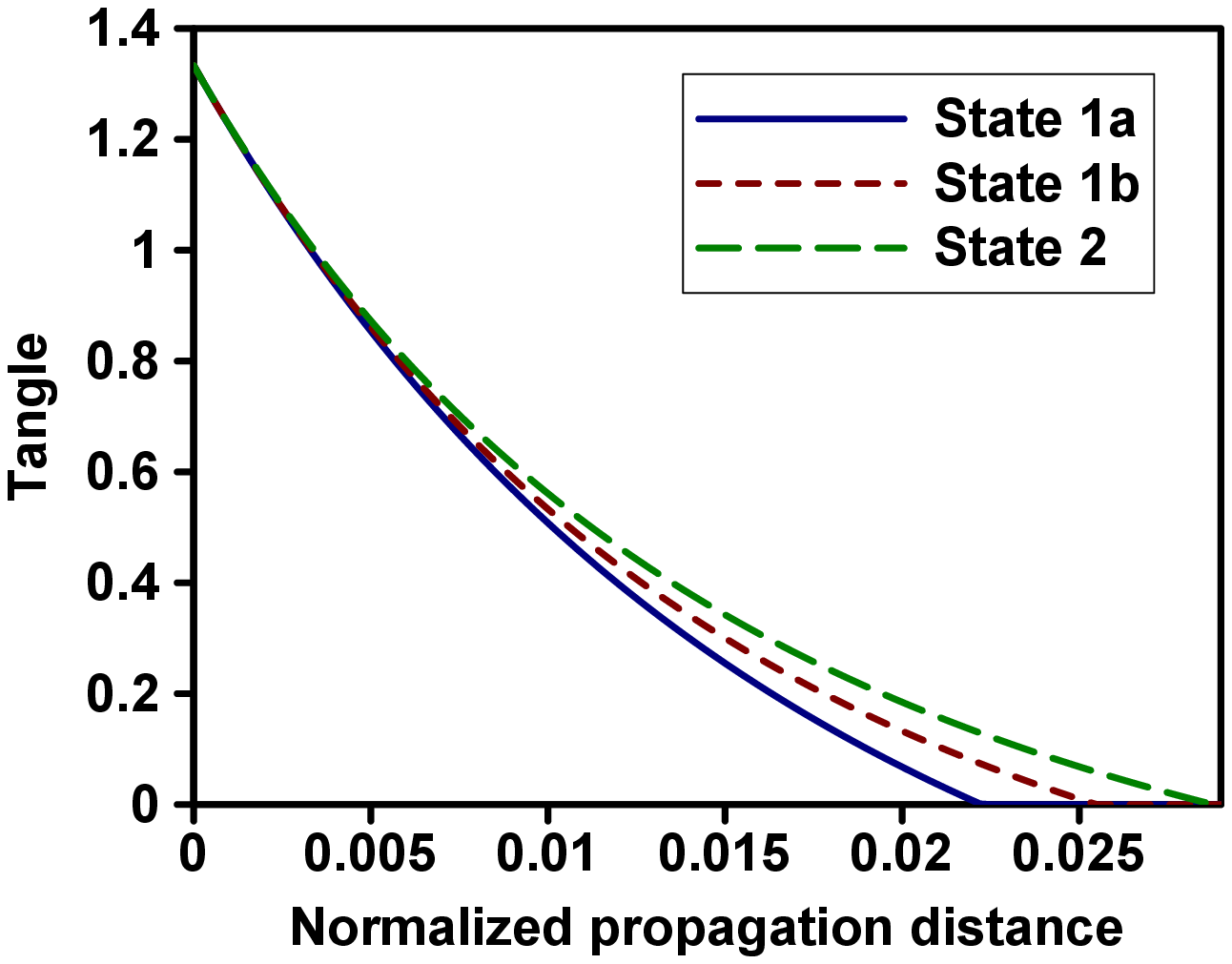}}\label{fig:me1}}
\subfloat[]{\scalebox{0.48}{\includegraphics{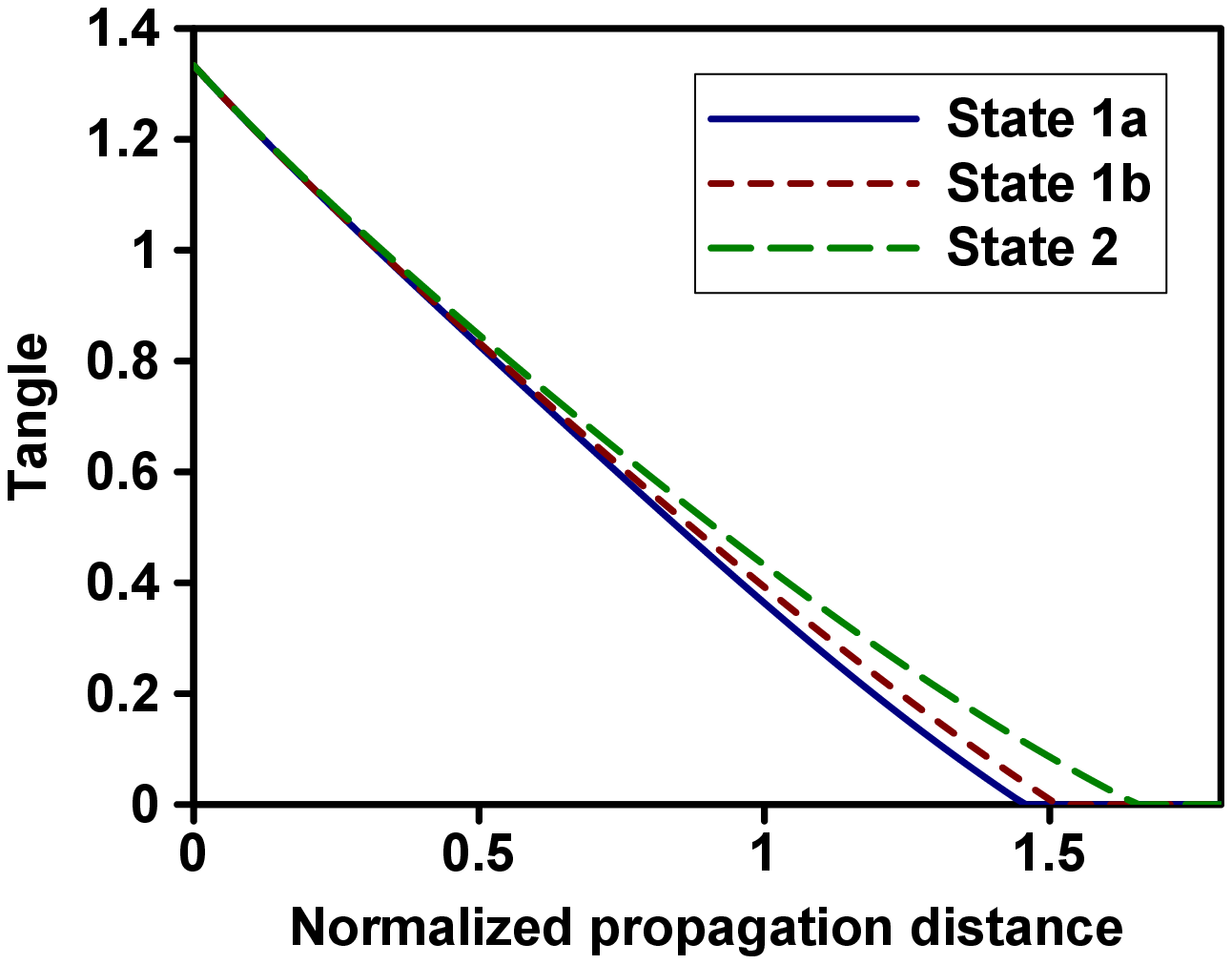}}\label{fig:me2}}
\end{center}
\caption{Tangle evolution of the initially maximally entangled state in strong turbulence (a) and in weak turbulence (b), both showing three curves: State 1a, State 1b and State 2. In strong turbulence State 1a and State 1b are obtained for $\phi=\pi$ and $\phi=0$, respectively, and in weak turbulence State 1a and State 1b are obtained for $\phi=0.4\pi$ and $\phi=-0.6\pi$, respectively.} 
\label{fig:me}
\end{figure}

The decay curves of State 1a, State 1b and State 2 are shown in figure \ref{fig:me}, both in weak and strong turbulence. One finds that states represented by State 2 are more robust than those represented by State 1. This again follows from the stronger coupling between differing magnitudes of azimuthal indices compared to the coupling strength between modes with identical magnitudes of azimuthal indices.

Comparing the curves for strong turbulence (figure \ref{fig:me1}) to the ones for weak turbulence (figure \ref{fig:me2}), one can see that the entanglement of states for larger propagation distances decays comparatively faster --- if we would have used the approximations for $Z(t)$, $H_{\rm r}(t)$ and $H_{\rm i}(t)$, given in (\ref{eq:Z}), (\ref{eq:Hr}) and (\ref{eq:Hi}), respectively, to calculate the tangle evolution for weak turbulence (i.e.\ when $\sigma = 0.25$), we would have found that the entanglement was maintained much longer than it does for the exact expressions. This can be seen from the fact that the tangle for weak turbulence decays nearly linearly all the way to zero entanglement, whereas for strong turbulence it decays linearly up to $t \approx 0.005$, followed by a nonlinear slower decay. Thus, if one can operate in a scenario where $\sigma$ and $z_{\rm R}$ have larger values, then one would have robust states for comparatively larger propagation distances.

\subsection{Most robust entangled state}

Upon comparing the curves for initially maximally entangled qutrit states in figure \ref{fig:me} to those for the Bell states of the two-dimensional subsystems in figure \ref{fig:Bell12}, we find that the maximally entangled qutrit states are not the most robust qutrit states. The Bell states of Set 3 maintain their entanglement longer than any of the initially maximally entangled qutrit states in (\ref{eq:state1}) and (\ref{eq:state2}), even though the initial entanglement of the Bell states is smaller. As a result, if one wants to find the most robust entangled state for a specific $t$, it is not enough to consider only those states that are initially maximally entangled.

\begin{figure}[!ht]
\begin{center}
\subfloat[]{\scalebox{0.48}{\includegraphics{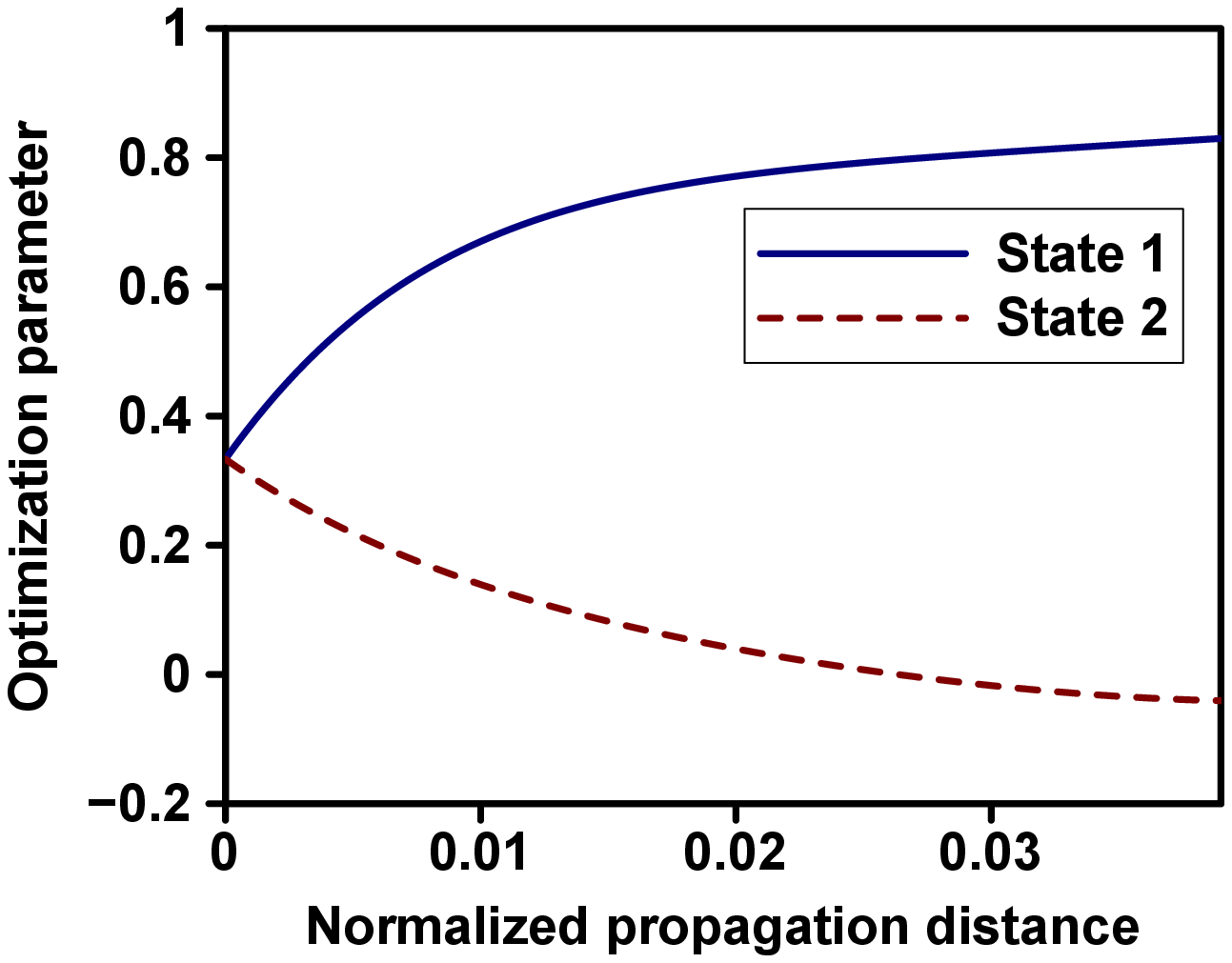}}\label{fig:q1}}
\subfloat[]{\scalebox{0.48}{\includegraphics{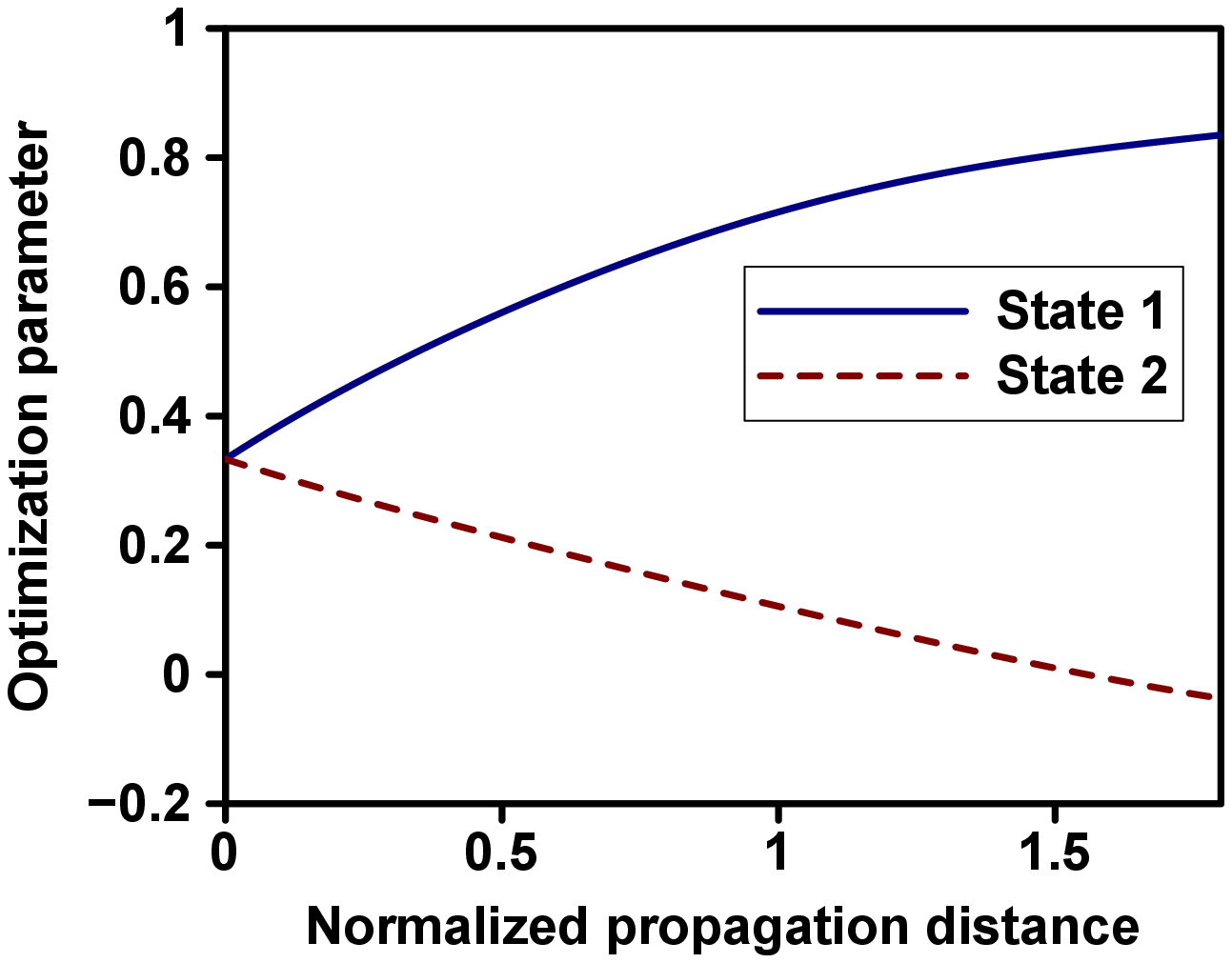}}\label{fig:q2}}
\end{center}
\caption{Optimization parameters $\mu_1(t)$ for State 1 and $\mu_2(t)$ for State 2 that gives the most robust entangled states $\ket{\mu_1(t)}$ and $\ket{\mu_2(t)}$ as a function of the normalized propagation distance $t$, in strong turbulence (a) and weak turbulence (b).} 
\label{fig:q12}
\end{figure}

To find the most robust states we optimize the tangle at a particular point $t=t_0$, for the general initial pure state, using the parameterization from (\ref{eq:parm}). This optimization process is similarly to \cite{mintert3}. From the optimization results, expressed in terms of the parameters $k_{\rm a}$, ..., $k_{\rm h}$ and $q_{\rm a}$, ..., $q_{\rm h}$, we found that all the states have one of the following two forms
\begin{eqnarray}
\ket{\mu_1} = & \frac{1}{\sqrt{2}} \cos(\mu_1) \exp(\rmi \phi_1) \ket{j,k} + \frac{1}{\sqrt{2}} \cos(\mu_1) \exp(\rmi \phi_2) \ket{-j,-k} \nonumber \\
& + \sin(\mu_1) \exp(\rmi \phi_3) \ket{0,0} \label{eq:q1} \\
\ket{\mu_2} = & \frac{1}{\sqrt{2}} \cos(\mu_2) \exp(\rmi \psi_1) \ket{0, -j} + \frac{1}{\sqrt{2}} \cos(\mu_2) \exp(\rmi \psi_2) \ket{-j, 0} \nonumber \\
& + \sin(\mu_2) \exp(\rmi \psi_3) \ket{j, j} , \label{eq:q2}
\end{eqnarray}
where we introduced the optimization parameters $\mu_1$ and $\mu_2$ in addition to the phases $\phi_1$, $\phi_2$, $\phi_3$, $\psi_1$, $\psi_2$ and $\psi_3$. These parameters are used to replace those given in (\ref{eq:parm}) for the sake of simplicity. The optimization parameters $\mu_1$ and $\mu_2$ determine the degree of relative weighting between the maximally entangled states in (\ref{eq:state1}) and (\ref{eq:state2}) and the Bell states of the two-dimensional subspaces given in (\ref{eq:set1}), (\ref{eq:set2}) and (\ref{eq:set3}).

\begin{figure}[!ht]
\begin{center}
\subfloat[]{\scalebox{0.48}{\includegraphics{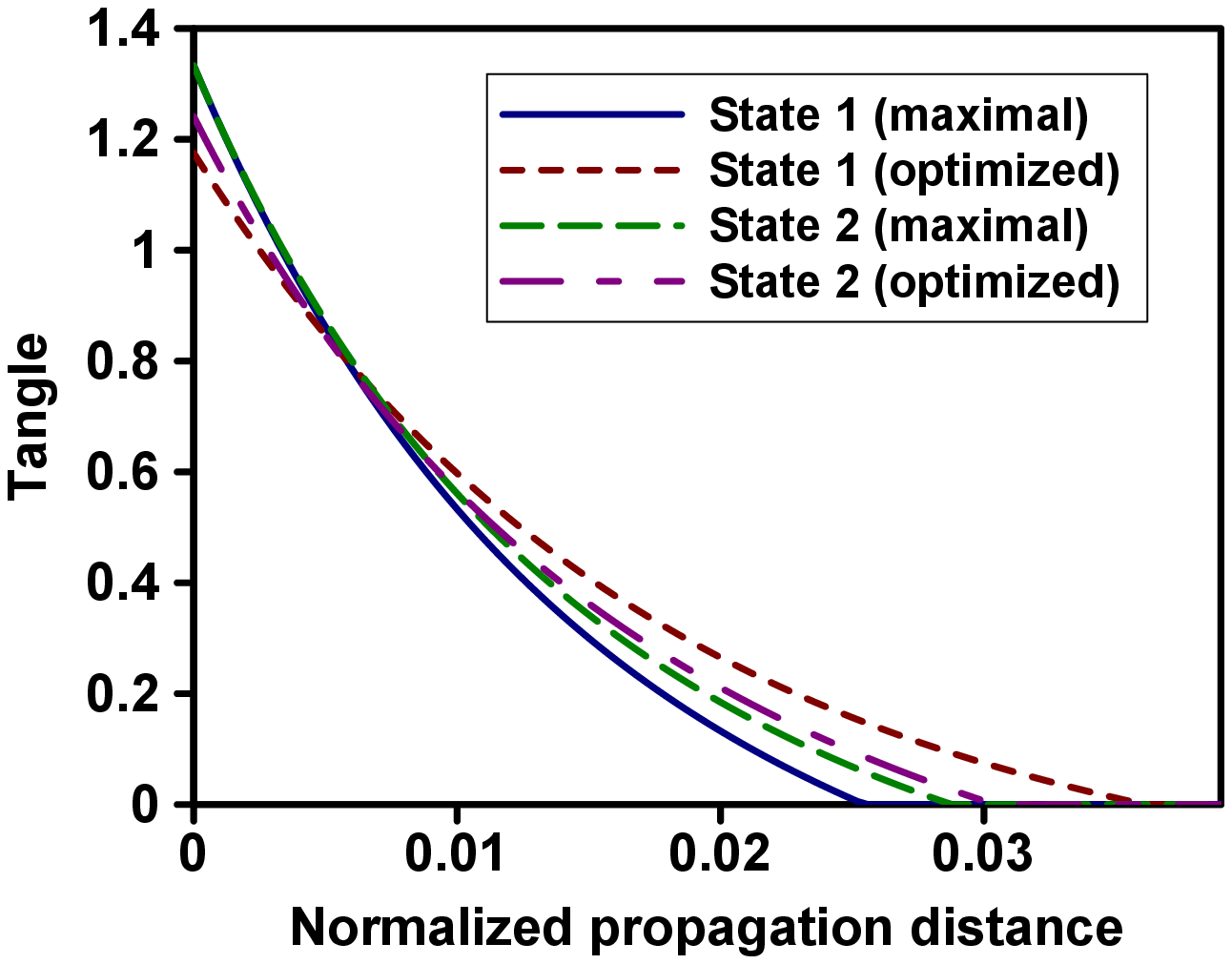}}\label{fig:opt12a}}
\subfloat[]{\scalebox{0.48}{\includegraphics{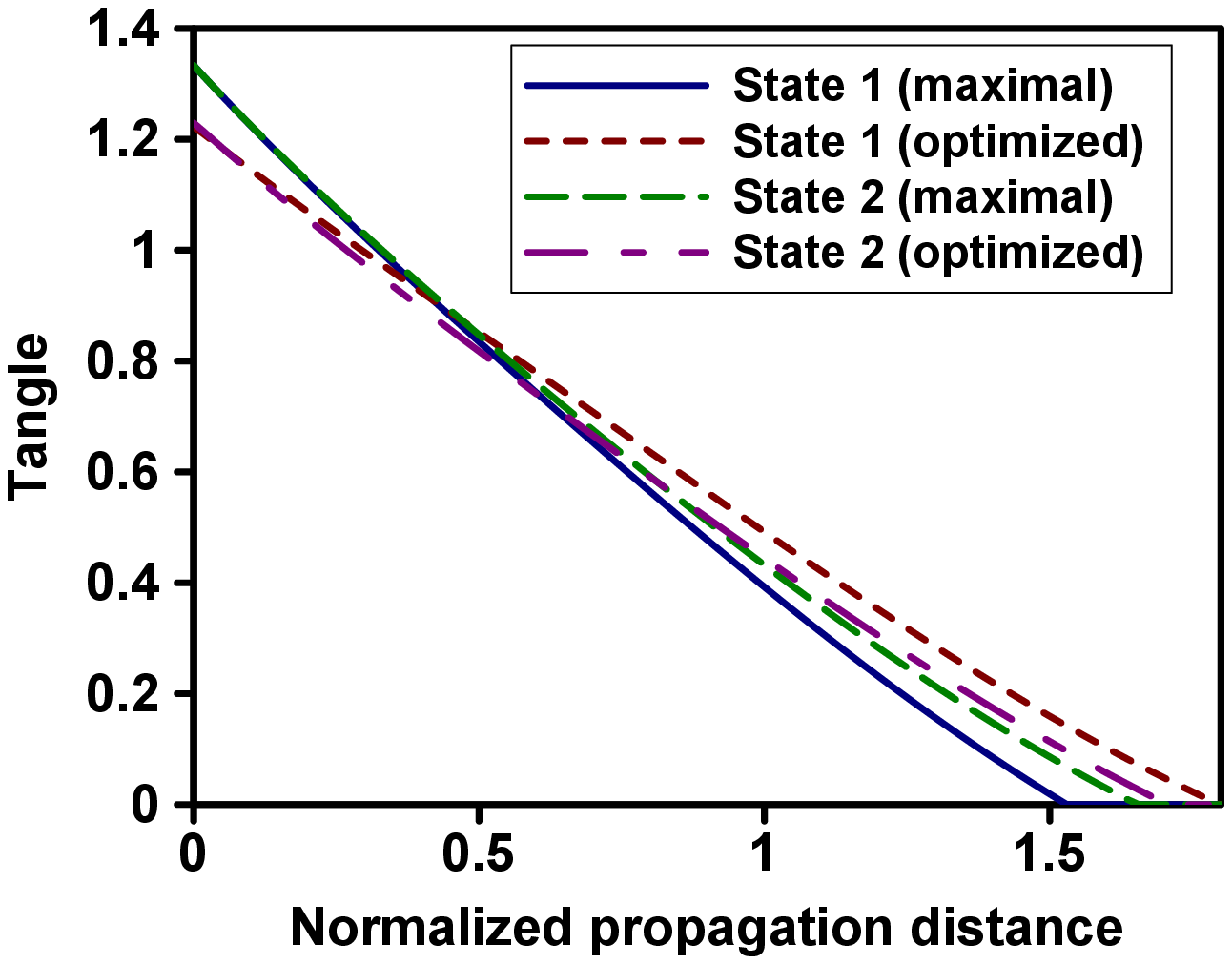}}\label{fig:opt12b}}
\end{center}
\caption{Comparison of the tangle evolution of initially maximally entangled states (State 1 and State 2) to those of the optimized states ($\ket{\mu_1(t)}$ and $\ket{\mu_2(t)}$) for strong turbulence optimized at $t_0=0.03$ (a) and for weak turbulence optimized at $t_0=1.6$ (b).} 
\label{fig:opt12}
\end{figure}

The curves of the optimized parameters $\mu_1(t)$ and $\mu_2(t)$ are shown in figure \ref{fig:q12}, for both weak and strong turbulence. The fact that the parameters are not independent of $t$ means that there does not exist one single state that is the most robust state for all propagation distances. Instead there are different states that have the highest remaining entanglement at each specific value of $t$. For $t=0$ the optimized states $\ket{\mu_1(0)}$ and $\ket{\mu_2(0)}$ are the maximally entangled states of (\ref{eq:state1}) and (\ref{eq:state2}), respectively.

Substituting (\ref{eq:phi}) and the optimal values for $\mu_1$ and $\mu_2$ at $t_0 = 0.03$ ($t_0 = 1.6$) for strong (weak) turbulence into (\ref{eq:q1}) and (\ref{eq:q2}), we obtain states that maintain a non-zero entanglement much longer than the initially maximally entangled states of (\ref{eq:state1}) and (\ref{eq:state2}) or the Bell states in (\ref{eq:set1}), (\ref{eq:set2}) and (\ref{eq:set3}). The corresponding curves of the tangle are shown in figure \ref{fig:opt12}. Note that State 1, which is less robust than State 2, can be optimized to produce state $\ket{\mu_1}$ that is more robust for large propagation distances than the optimized state $\ket{\mu_2}$. It turns out that, when optimized, $\ket{\mu_1}$ is the most robust qutrit state within our three-dimensional Hilbert space for propagation through turbulence.

\section{Summary and conclusions}

We have shown that for three-dimensional systems, the initial amount of entanglement does not uniquely define the amount of entanglement left after a non-zero propagation through a turbulent medium. This highlights the fact that the upper bound for the entanglement evolution \cite{tiersch1, tiersch2} has to be treated with caution. The reason is that in our three-dimensional Hilbert space the coupling strength between differing magnitudes of azimuthal indices is different from the coupling strength between identical magnitudes. This follows from our choice of Hilbert space and the fact that the coupling strength decreases as the difference between the azimuthal modal indices increases.

By optimizing the tangle, starting from the most general initial pure state, we found that neither the initially maximally entangled states, nor one of the Bell states of the two-dimensional subsystems are the most robust states. Instead it is a coherent superposition between both. For each propagation distance one can find a specific initial entangled state that retains the most entanglement up to that propagation distance. Thus, there is not one single most robust state for all distances.

The effect of the optimization of the robustness of the qutrit states is still rather small (it only increases the propagation distance by a few percent). However, it is reasonable to expect that the effect of such an optimization would become more significant for higher dimensional Hilbert spaces, making it worth doing the optimization to find robust quantum states in higher dimensional Hilbert spaces and thereby improve the performance of quantum communication systems. As a result we believe that the optimization of the robustness of higher dimensional quantum states will have a significant impact in practical free-space quantum communication systems.

\ack

We gratefully acknowledge discussions with Andreas Buchleitner. This work was done with funding support from the CSIR.

%\References

\section*{References}

%\bibliographystyle{unsrt}
%\bibliography{robustpub}

\end{document}